\documentstyle[rotate,dina4,psfig]{article}

\def\IJMPA{{\em Int. J. Mod. Phys. A}}
\def\IJMPC{{\em Int. J. Mod. Phys. C}}

\def\NPB{{\em Nucl. Phys. B}}
\def\PLB{{\em Phys. Lett.  B}}
\def\PRL{{\em Phys. Rev. Lett.}}

\def\PRD{{\em Phys. Rev. D}}

\def\HH{6.0cm}          
\def\WD{5.0cm}          
\def\VS{1.5cm}          

\newcommand{\be}{\begin{equation}}
\newcommand{\ee}{\end{equation}}
\newcommand{\ba}{\begin{eqnarray}}
\newcommand{\ea}{\end{eqnarray}}
\newcommand{\NL}{\nonumber \\ }
\newcommand{\ra}{\rightarrow}
\def\vep{\varepsilon}

\begin{document}
\title{{\bf Chaotic behavior of confining lattice \\
     gauge field configurations }} 
\author{
{\sc Tam\'as S. Bir\'o$^1$, Markus Feurstein$^{2}$ 
and Harald Markum$^{2}$ }\\[2.812mm]
\normalsize \hspace*{-8pt}$^1$   
Research Institute for Particle and Nuclear Physics\\ 
H-1525 Budapest, Pf.49, Hungary\\[0.2ex] 
\hspace*{-8pt}$^2$ Institute for Nuclear Physics, TU-Vienna\\ 
A-1040 Wien, Wiedner Hauptstrasse 8-10, Austria  
 }
\maketitle
\begin{abstract} 
 	We analyze the leading Lyapunov exponents of 
  	SU(2) Yang-Mills field configurations on the lattice
        which are initialized by quantum Monte Carlo simulations.
        We find that configurations in the strong coupling phase
        at finite temperature are substantially more chaotic
        than in deconfinement.
\end{abstract}


\setcounter{page}{1}
 
\section{Motivation}

The study of chaotic dynamics of classical field
configurations in field theory finds its motivation
in two main areas: i) particle production from
long wavelength, collective fields
and ii) thermalization
due to entropy production in the equipartioning
process.
Chaos is known to lead to mode sharing, i.e. the
energy concentrated initially at a few selected
wavelengths tends to be shared among all modes.
The hard ones of these show a dispersion relation
close to that of free particles \cite{BOOK}.

In such processes the initial configurations are usually described
by nonperturbative methods, while the final state is
characterized by perturbative physics.
Examples are the baryon production at the electroweak
phase tran\-sition \cite{THIS.CONF},  enhanced pion production due
to disoriented chiral condensate formation in energetic
hadronic collisions \cite{DCC.PEOPLE} or thermalized quark-gluon matter
production from color ropes in relativistic heavy ion
collisions \cite{COLOR.ROPE},\cite{CHAOS.PRL}.
The chaotic dynamics also can be used for
preparing thermal initial conditions for a simulation.


The numerical simulation of the gauge field dynamics
is usually done on a lattice, because this method
ensures an adequate treatment of gauge symmetries
and is non-perturbative \cite{WILSON},\cite{CREUTZ}.
Special care is
taken for satisfying constraints, especially Gauss law,
by the Hamiltonian simulation of the lattice dynamics \cite{KOGUT}.
Algorithmically also the correct implementation of the Gauss law is
important. It is achieved by using the Noether charge
corresponding to the finite time step recursion form
of the equations of motion \cite{ALGO.IJMP}.

Besides the above mentioned areas of application
a pure theoretical motivation drives the study of
chaotic field dynamics as well: its possible relation to
the confinement problem is a longstanding question \cite{MATYNIAN},\cite{SAVVIDY}.
Both confinement and chaotic behavior are related
to long wavelength field components, so already
in the earliest simple models this question has been
raised \cite{XY-MODEL},\cite{SPAGETTI}.

Recently a discussion was initiated \cite{NIELSEN96},\cite{SMIT}
about the continuum limit of the classical lattice gauge theory.
It was conjectured that for small energy per plaquette the lattice
system would break the linear scaling between Lyapunov
exponents and energy. This finding was shown to possibly be
a finite time artifact \cite{MULLER96} and the linear
correspondence was reestablished.
Nevertheless it could not be decided whether inspite the divergence
of both the energy and Lyapunov exponent in the continuum
limit (lattice spacing zero) their ratio would scale.
Our method of preparing the initial field configurations
by quantum Monte Carlo simulations may shed some light
on this question, because this way an explicit
coupling dependence occurs in the results.
Using the perturbative $\beta$-function the
continuum limit of the above mentioned ratio can be
considered.

In this paper we also search for a possible connection
between chaotic dynamics and confinement in Yang-Mills
theory by numerical simulation. We do this by investigating
the chaotic dynamics of SU(2) Yang-Mills fields on
a three dimensional lattice. The starting configurations
are prepared by four dimensional quantum
Monte Carlo techniques for finite temperature QCD undergoing
a phase transition to quark-gluon plasma. 
This way we expect to see
a coincidence between the strong coupling phase and the strength of
chaotic behavior in lattice simulations.

After reviewing essential
definitions of the physical quantities describing chaos
and their computation in lattice gauge theory \cite{BOOK},\cite{CHAOS.PRL}
we outline our method for the extraction of starting
configurations of a three dimensional Hamiltonian dynamics  from
four dimensional euclidean field configurations.
Our results are then presented by showing an example of the
exponential divergence of small initial distances between
nearby field configurations. It is followed by a detailed study
of the maximal Lyapunov exponent and average plaquette energy
as function of the coupling strength.


\section{Classical chaotic dynamics }

Chaotic dynamics in general is characterized by the
spectrum of Lyapunov exponents. These exponents, if they are positive,
reflect an exponential divergence of initially adjacent configurations.
In case of symmetries inherent in the Hamiltonian of the system
there are corresponding zero values of these exponents. Finally
negative exponents belong to irrelevant directions in the phase
space: perturbation components in these directions die out
exponentially. Pure gauge fields on the lattice show a characteristic
Lyapunov spectrum consisting of one third of each kind of
exponents \cite{BOOK},\cite{CHAOS.IJMP}.
This fact reflects the elimination of
longitudinal degrees of freedom of the gauge bosons.
Assuming this general structure of the Lyapunov spectrum we
investigate presently its magnitude only, namely the maximal
value of the Lyapunov exponent, $L_{{\rm max}}$.

The general definition of the Lyapunov exponent is based on a
distance measure $d(t)$ in phase space,
\be
L := \lim_{t\ra\infty} \lim_{d(0)\ra 0}
\frac{1}{t} \ln \frac{d(t)}{d(0)}.
\ee
In case of conservative dynamics the sum of all Lyapunov exponents
is zero according to Liouville's theorem,
\be
\sum L_i = 0.
\ee
We utilize the gauge invariant distance measure consisting of
the local differences of energy densities between two field configurations
on the lattice:
\be
d : = \frac{1}{N_P} \sum_P\nolimits \, \left| {\rm tr} U_P - {\rm tr} U'_P \right|.
\ee
Here the symbol $\sum_P$ stands for the sum over all $N_P$ plaquettes,
so this distance is bound in the interval $(0,2N)$ for the group
SU(N). $U_P$ and $U'_P$ are the familiar plaquette variables, constructed from
the basic link variables $U_{x,i}$,
\be
U_{x,i} = \exp \left( aA_{x,i}^cT^c \right),
\ee
located on lattice links pointing from the position $x=(x_1,x_2,x_3)$ to
$x+ae_i$. The generators of the group are
$T^c = -ig\tau^c/2$ with $\tau^c$ being the Pauli matrices
in case of SU(2) and $A_{x,i}^c$ is the vector potential.
The elementary plaquette variable is constructed for a plaquette with a
corner at $x$ and lying in the $ij$-plane as
\be
U_{x,ij} = U_{x,i} U_{x+i,j} U^{\dag}_{x+j,i} U^{\dag}_{x,j}.
\ee
It is related to the magnetic field strength $B_{x,k}^c$:
\be
U_{x,ij} = \exp \left( \vep_{ijk} a B_{x,k}^c T^c \right).
\ee
The electric field strength $E_{x,i}^c$ is related to the canonically conjugate
momentum $P_{x,i} = \dot{U}_{x,i}$ via
\be
E^c_{x,i} = \frac{2a}{g^3} {\rm tr} \left( T^c \dot{U}_{x,i} U^{\dag}_{x,i} \right).
\ee

The spectrum of Lyapunov exponents is  connected to the Kolmogorov-Sinai
entropy
\be
\sigma_{KS} = \sum_{L_i > 0 } L_{i}.
\label{KS}
\ee
This quantity describes the rate by which information is lost during the
dynamical evolution of initially adjacent phase space points if the resolution
is kept fixed (e.g. the smallest phase space cell has a size of the
Planck constant, $h=2\pi$). 
This information loss can also be regarded as the entropy generation 
rate. Dividing the sum (\ref{KS}) by the number of
Lyapunov exponents (proportional to the lattice volume), we are
approximating the continuum entropy density generation rate for big
lattices. Calculations on very big lattices have, of course, their limits
in the available computer resources.

Eventually the initial energy per degree of freedom, the lattice approximation
to the energy density, can be related to a temperature $T$ via the classical
equipartition theorem. In fact, whenever the dynamics has positive Lyapunov
exponents, equipartition is guaranteed. In this way, if the equation of state
in the final state is known, a characteristic time scale can be obtained:
\begin{itemize}
\item the equation of state determines the entropy $S(E)$,
\item the equipartition leads to $T(E)$ and finally
\item the Kolmogorov entropy yields $\dot{S}(E)$.
\end{itemize}
The characteristic time (inverse thermalization rate) is then obtained as
\be
t_T = S / \dot{S}.
\ee
Furthermore, preparing the initial state by quantum Monte Carlo, a correspondence
between the coupling strength and the Lyapunov exponents can be obtained.
This can eventually identify the entropy generation potential of given
field configurations and explore possible coincidences between confinement
in lattice gauge theory and chaotic classical dynamics. This is the
aim of our present study.


\section{Initial states prepared by quantum Monte Carlo }

The Hamiltonian of the lattice gauge field system can be casted into
the form
\be
H = \sum \left[ \frac{1}{2} \langle P, P \rangle \, + \,
 1 - \frac{1}{4} \langle U, V \rangle \right].
\ee
Here the scalar product between group elements stands for
$\langle A, B \rangle = \frac{1}{2} {\rm tr} (A B^{\dag} )$.
The staple variable $V$ is a sum of triple products of elementary
link variables closing a plaquette with the chosen link $U$.
This way the Hamiltonian is formally written as a sum over link
contributions and $V$ plays the role of the classical force
acting on the link variable $U$. The naive equations of motion
following from this Hamiltonian, however, have to be completed
in order to fulfill the constraints
\ba
\langle U, U \rangle &=& 1, \NL
\langle P, U \rangle &=& 0.
\ea
The following finite time step recursion
formula:
\ba
U' &=& U + dt ( P' - \vep U ), \NL
P' &=& P + dt ( V - \mu U + \vep P' ),
\ea
with the Lagrange multipliers
\ba
\vep &=& \langle U, P' \rangle, \NL
\mu &=& \langle U, V \rangle + \langle P', P' \rangle,
\ea
conserves the Noether charge belonging to the Gauss law,
\be
\Gamma = \sum_+ PU^{\dag} - \sum_- U^{\dag}P.
\ee
Here the sums indicated by $+$ run over links starting from,
and those by $-$ ending at a given
site $x$, where the Noether charge $\Gamma$ is defined.
The above algorithm is written in an implicit form, but it can be
casted into explicit steps, so no iteration is necessary \cite{ALGO.IJMP}.

Initial conditions chosen randomly with a given average magnetic energy
per plaquette have been investigated in past years. A linear scaling
of the maximal Lyapunov exponent with the total energy of the system has been
established for different lattice sizes and coupling strengths \cite{BOOK},\cite{CHAOS.PRL}.
In the present study we prepare the initial field configurations
from a standard four dimensional euclidean Monte Carlo program on
a $12^3\times 4$ lattice varying $\beta = 4/g^2$.

We relate such four dimensional euclidean
lattice field configurations to minkow\-skian momenta and fields
for the three dimensional Hamiltonian simulation
by the following approach:

First we fix a time slice of the four dimensional lattice.
We denote the link variables in the three dimensional sub-lattice
by $U' = U_i(x,t).$ Then we build triple products on attached handles
in the positive time direction, \newline
\hbox{$U'' = U_4(x,t)U_i(x,t+a)U^{\dag}_4(x+a,t)$.}
We obtain the canonical variables of the Hamiltonian system
by using
\ba
P &=& (U'' - U') / dt, \NL
U &\propto & (U'' + U').
\ea
Finally $U$ is normalized to $\langle U, U \rangle = 1$.

This definition constructs the momenta according to a
simple definition of the timelike covariant derivative.
The multiplication with the link variables
in time direction can also be viewed as a gauge transformation to
$U_4(x,t)=1$, i.e. $A_0=0$ Hamiltonian gauge.


\section{Chaos and confinement  }

The review of our results we start with a characteristic example
of the time evolution of the distance between initially adjacent
configurations. An initial state prepared by a standard four dimensional
Monte Carlo simulation is evolved according to the classical Hamiltonian dynamics
in real time. Afterwards this initial state is rotated locally by
group elements which are chosen randomly near to the unity.
The time evolution of this slightly rotated configuration is then
pursued and finally the distance between these two evolutions
is calculated at the corresponding times.
A typical exponential rise of this distance followed by a saturation
can be inspected in Fig. 1. While the saturation is an artifact of
the compact distance measure of the lattice, the exponential rise
(the linear rise of the logarithm)
can be used for the determination of the leading Lyapunov exponent.
The naive determination and more sophisticated rescaling methods lead to
the same result.
Deviations have been observed only for very small energies \cite{MULLER}.


\begin{figure}[h]
\rule{5cm}{0.2mm}\hfill\rule{5cm}{0.2mm}
\vskip \VS
\centerline{\rotate[r]{\psfig{figure=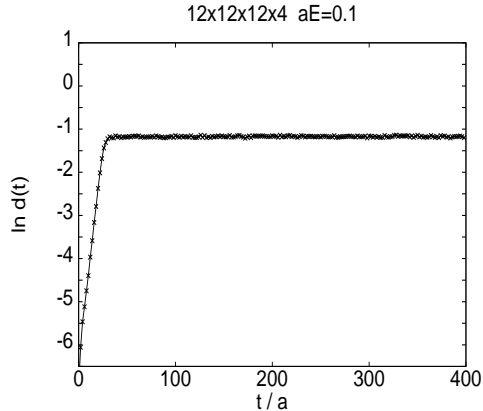,height=\HH,width=\WD}}}
\vspace{0.3cm}
\caption[Fig1]{
  Exponentially diverging distance of initially adjacent field
  configurations on a $12^3$ lattice prepared at $4/g^2 = 2.3$
  and $g^2aE = 0.175$.
\label{Fig1}
 }

\end{figure}


The main result of the present study is the dependence of the leading
Lyapunov exponent $L_{{\rm max}}$ on the inverse coupling strength $4/g^2$
displayed in Fig. 2.
As expected the strong coupling phase, where confinement
of static quarks
has been established many years ago by proving the area law
behavior for large Wilson loops, is more chaotic.
The transition reflects the critical temperature to the
deconfinement phase.
Furthermore the maximal Lyapunov exponent varies between the
strong coupling and weak coupling regime more pronounced
than the average energy per plaquette.
Fig. 3 shows the somewhat softer transition of the
energy per plaquette as a function of the inverse coupling strength.


\begin{figure}[h]
\rule{5cm}{0.2mm}\hfill\rule{5cm}{0.2mm}
\vskip \VS
\centerline{\rotate[r]{\psfig{figure=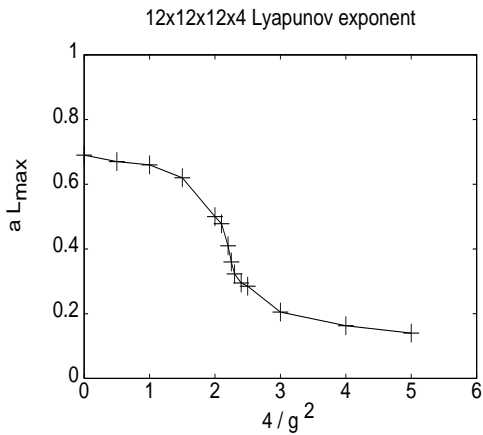,height=\HH,width=\WD}}}
\vspace{0.3cm}
\caption[Fig2]{
  Maximal Lyapunov exponent in lattice units as a function
  of the inverse coupling strength $\beta = 4/g^2$.
 }

\end{figure}


Although in the transition region the behavior of the maximal
Lyapunov exponent does not follow exactly that of the average
energy, the qualitative correspondence is clearly seen.
We compare therefore maximal Lyapunov exponents stemming from
classical random gauge fields in the initial state, where a linear scaling
between the leading exponent and the total energy has been
established.
Fig. 4 shows the energy dependence of the Lyapunov exponents for both choices.
This result agrees nicely with the physical picture of random
fields being the background for confinement and should be
an encouragement for random matrix calculations of the QCD
vacuum.


\begin{figure}[h]

\rule{5cm}{0.2mm}\hfill\rule{5cm}{0.2mm}
\vskip \VS
\centerline{\rotate[r]{\psfig{figure=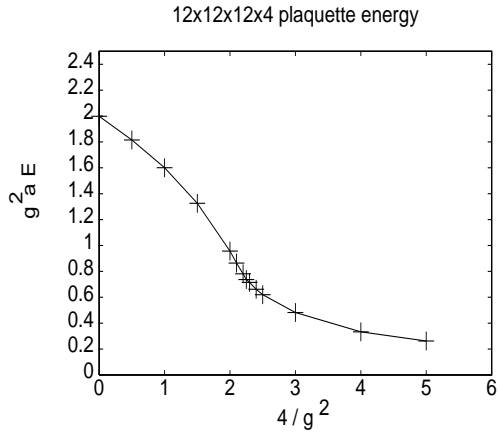,height=\HH,width=\WD}}}
\vspace{0.3cm}
\caption[Fig3]{
  Transition of the average plaquette energy
  as a function of the inverse coupling strength $\beta = 4/g^2$.
 }

\end{figure}



\begin{figure}[h]

\rule{5cm}{0.2mm}\hfill\rule{5cm}{0.2mm}
\vskip \VS
\centerline{\rotate[r]{\psfig{figure=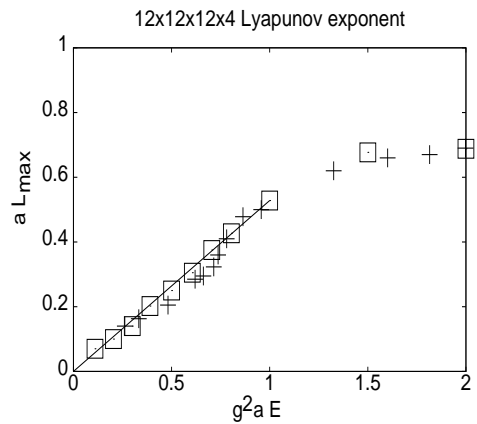,height=\HH,width=\WD}}}
\vspace{0.3cm}
\caption[Fig4]{
  Comparison of maximal Lyapunov exponents for initial states
  prepared by quantum Monte Carlo simulation (crosses) and by classical
  randomization (squares) as a function of the scaled energy per
  plaquette $ag^2E$.
 }

\end{figure}


A comparison with the scaling law of random configurations 
is made in Fig. 5
by plotting the linear scaling parameter $L_{{\rm max}}/(g^2E)$
versus the degree of equipartition $E/T$. Here 
the temperature $T$ is taken from
the euclidean lattice size $N_t=4$ via $1/T=N_ta$ to obtain dimensionless
quantities.
There seem to be systematic deviations for
the  Lyapunov exponent from the quantum Monte Carlo procedure.
The results presented here correspond to an
exemplatory study, albeit the time evolution itself
covers the available phase space ergodically.


\begin{figure}[h]

\rule{5cm}{0.2mm}\hfill\rule{5cm}{0.2mm}
\vskip \VS
\centerline{\rotate[r]{\psfig{figure=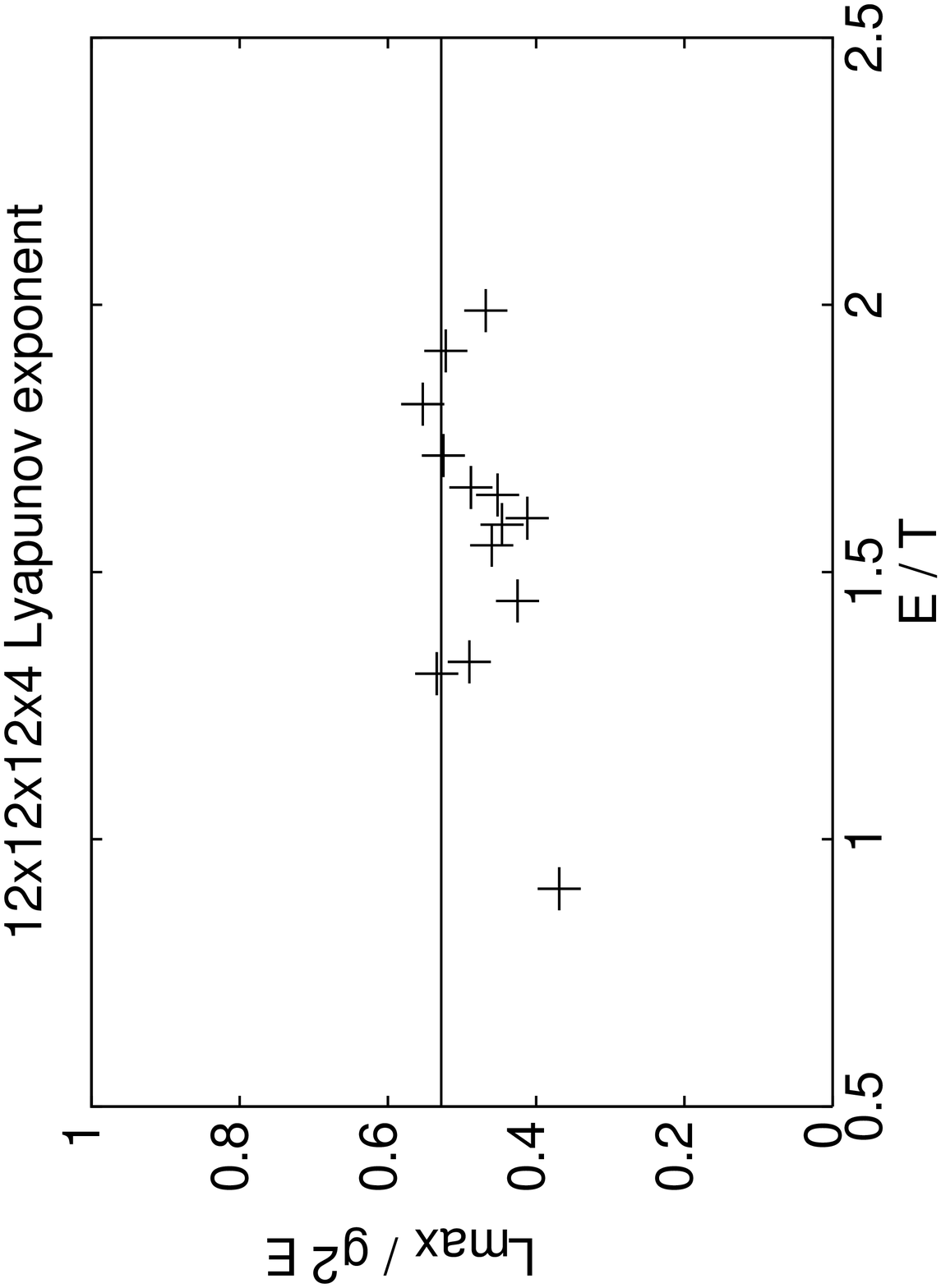,height=\HH,width=\WD}}}
\vspace{0.3cm}
\caption[Fig5]{
  Scaling of the Lyapunov exponent, $L_{{\rm max}}/(g^2E)$,
  as a function of the effective physical degrees of freedom,
  $E/T$. The line corresponds to the classical scaling law.
 }

\end{figure}

Finally it is of interest whether a continuum limit
exists for the Lyapunov exponent. Although it is expected to diverge
like $1/a$ as the lattice spacing vanishes, its ratio with
the average energy per plaquette does not seem to show
dramatic changes in Fig. 6. The inverse coupling $4/g^2$
and the lattice spacing are connected by the 
renormalization group formula for SU(2).
One should be aware that the original euclidean system
undergoes a phase transition and the small $a$ limit
lies in the quark-gluon plasma phase.
This gives hope that {\em some} ratios of classically
divergent quantities still might be extrapolated to the
continuum theory.


\begin{figure}[t]

\rule{5cm}{0.2mm}\hfill\rule{5cm}{0.2mm}
\vskip \VS
\centerline{\rotate[r]{\psfig{figure=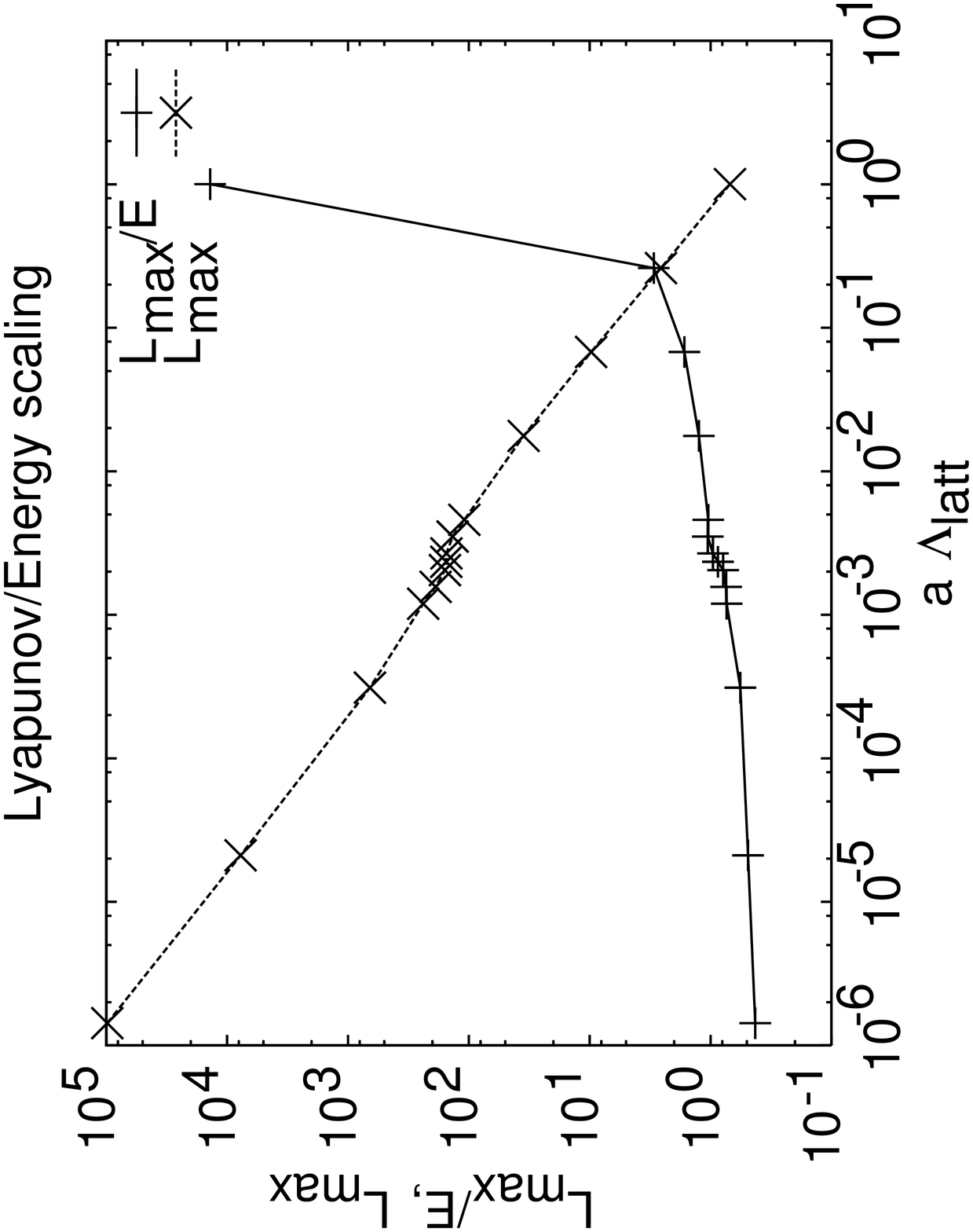,height=\HH,width=\WD}}}
\vspace{0.3cm}
\caption[Fig6]{
  Scaling of the Lyapunov exponent, $L_{{\rm max}}$,
  and the ratio $L_{{\rm max}}/E$ as a function of the lattice spacing.
 }

\end{figure}



Summarizing we investigated the classical chaotic dynamics of
SU(2) lattice gauge field configurations prepared by finite
temperature quantum Monte Carlo simulation. We established
that chaos develops as in case of the random choice
of initial configurations. The maximal Lyapunov exponent
shows a sharp transition as a function of the coupling strength:
the strong coupling (confining) phase is substantially more chaotic.
The results are close to that of random initial
states suggesting that the random matrix model of confinement
is a good approximation also for the real time classical
dynamics. Finally the scaling law between the leading
Lyapunov exponent and energy per plaquette seems to hold
when extrapolating to the continuum limit at finite
temperature.
Whether this is also true at zero temperature, investigations
with configurations initialized on four dimensional symmetric
lattices may answer.


\section*{ Acknowledgements  }

This work has been supported by the Aktion \"Osterreich-Ungarn
under the project 25\"o1 and by the Hungarian National
Scientific Fund under the project OTKA T019700.
Discussions with W. Sakuler are gratefully acknowledged.



\vfill\eject



\newpage
\begin{thebibliography}{COLOROROPEOO}

\bibitem{BOOK}  T.S.Bir\'o, S.G.Matinyan and B.M\"uller:
        {\em Chaos and Gauge Field Theory,} World Scientific,
        Singapore, 1995.

\bibitem{THIS.CONF} Proc. of the Workshop on Strong and Electroweak 
	Matter '97, May 21-25 1997, Eger, Hungary,
	World Scientific, Singapore, 1998.

\bibitem{DCC.PEOPLE}  	A.Anselm, \PLB {\bf 217}, 169 (1989),
			J.D.Bjorken, \IJMPA {\bf 7}, 4189 (1992),
			J.P.Blaizot and A.Krzywicki, \PRD {\bf 46}, 246 (1992).

\bibitem{COLOR.ROPE} 	T.S.Bir\'o, J.Knoll and H.B.Nielsen, \NPB {\bf 245}, 449
(1984).

\bibitem{CHAOS.PRL} 	B.M\"uller and A.Trayanov, \PRL \, {\bf 68}, 3387 (1992).

\bibitem{WILSON} 	K.G.Wilson, \PRD {\bf 10}, 2445 (1975).

\bibitem{CREUTZ} 	M.Creutz: {\em Quarks, Gluons and Lattices,}
        		Cambridge University Press, 1983.

\bibitem{KOGUT} 	J.Kogut and L.Susskind, \PRD {\bf 11}, 395 (1975).

\bibitem{ALGO.IJMP} 	T.S.Bir\'o, \IJMPC {\bf 6}, 327 (1995).

\bibitem{MATYNIAN}      S.G.Matinyan and G.Savvidy, \NPB {\bf 134}, 539
			(1978).
\bibitem{SAVVIDY}       S.K.Savvidy, \NPB {\bf 246}, 302 (1984).

\newpage
\bibitem{XY-MODEL}      S.Matinyan, G.Savvidy and N.Ter-Arutyunyan-Savvidy,
			{\em JETP Lett.} {\bf 34}, 590 (1981).

\bibitem{SPAGETTI}      H.Nielsen and P.Olesen, \NPB {\bf 61}, 45 (1973).


\bibitem{NIELSEN96}     H.B.Nielsen, H.H.Rugh and S.E.Rugh, chao-dyn/9605013,
			hep-th/9611128.

\bibitem{SMIT}          J.Smit, hep-lat/9710026.

\bibitem{MULLER96}      B.M\"uller, chao-dyn/9607001.

\bibitem{CHAOS.IJMP}    T.S.Bir\'o, C.Gong, B.M\"uller and A.Trayanov, 
			\IJMPC {\bf 5}, 113 (1994).

\bibitem{MULLER}        B.M\"uller, {\em private communication}, 1996.

\end{thebibliography}
\end{document}